\documentclass[aps,prl,notitlepage,twocolumn]{revtex4}

\usepackage{amssymb}
\usepackage{amsmath}
\usepackage{graphicx}
\usepackage{bm}
\usepackage{mdframed}
\usepackage{color}
\usepackage[mathscr]{euscript}
\usepackage{verbatim}
\usepackage{stmaryrd}
\usepackage[colorlinks=true]{hyperref}

\begin{document}

\bibliographystyle{prsty}

\title{Hydrodynamics of three-dimensional skyrmions in frustrated magnets}

\author{Ricardo Zarzuela}
\author{H\'{e}ctor Ochoa}
\altaffiliation[Current address: ]{Department of Physics, Columbia University, New York, NY 10027, USA}
\author{Yaroslav Tserkovnyak}

\affiliation{Department of Physics and Astronomy, University of California, Los Angeles, California 90095, USA}

\begin{abstract}
We study the nucleation and collective dynamics of Shankar skyrmions [R. Shankar, Journal de Physique \textbf{38}, 1405 (1977)] in the class of frustrated magnetic systems described by an SO(3) order parameter, including multi-lattice antiferromagnets and amorphous magnets. We infer the expression for the spin-transfer torque that injects skyrmion charge into the system and the Onsager-reciprocal pumping force that enables its detection by electrical means. The thermally-assisted flow of topological charge gives rise to an algebraically decaying drag signal in nonlocal transport measurements. We contrast our findings to analogous effects mediated by spin supercurrents.
\end{abstract}
\maketitle

{\it Introduction.}|The recent years have witnessed a growing interest in the transport properties of frustrated (quantum) magnets \cite{Yamashita-NPhys2009,Hirschberger-Sci2015,Hirschberger-PRL2015,Watanabe-PNAS2016,Zink-NPhys2017,Kasahara-Nat2018,OZ-PRB2018,Ye-2018,Hentrich-2018} since they provide a powerful knob to explore unconventional spin excitations and reveal the emergence of phases characterized by a highly degenerate ground state. Spin glasses \cite{SG,RMP-SG}, spin ices \cite{SPI} and spin liquids \cite{SPL}, to mention a few examples, belong to this broad family. In the exchange-dominated limit for magnetic interactions \cite{FN1}, long-wavelength excitations around a local free-energy basin (a classical ground-state manifold) are generically described by the O(4) nonlinear $\sigma-$model \cite{Azaria-PRL1992,Chubukov-PRL1994}
\begin{align}
\label{eq1}
 \hspace{-0.2cm}S=\frac{1}{4}\hspace{-0.03cm}\int \hspace{-0.05cm}d^{3}\vec{r}\,dt \hspace{-0.02cm}\left( \hspace{-0.02cm}\chi\,\text{Tr} \hspace{-0.04cm}\left[\partial_t\hat{R}^T\partial_t\hat{R}\right]\hspace{-0.07cm}- \hspace{-0.04cm}\mathcal{A}\,\text{Tr} \hspace{-0.04cm}\left[\partial_k \hat{R}^T\partial_k\hat{R}\right] \hspace{-0.02cm}\right) \hspace{-0.02cm},
\end{align}
where the order parameter $\hat{R}(\vec{r},t)$ represents smooth and slowly varying proper rotations of the initial noncoplanar spin configuration \cite{Halperin-PRB77,Dombre-PRB89}; $\chi$ and $\mathcal{A}$ denote the spin susceptibility and the order-parameter stiffness of the system, respectively. Phase-coherent precessional states sustain spin supercurrents \cite{OZ-PRB2018} that manifest themselves as a long-range spin signal decaying algebraically with the propagation distance. This form of spin superfluidity \cite{spin_sf} gives rise to a low-dissipation channel for spin transport that could be probed via nonlocal magnetotransport measurements \cite{vanWees}.

The SO(3) order parameter can also host stable three-dimensional solitons akin to skyrmions in chiral models of mesons \cite{Skyrme-NucPhys62}. In condensed matter physics these textures are known as Shankar skyrmions and appear, e.g., in the A-phase of superfluid $^3$He \cite{Shankar,Volovik_Mineev} and in atomic Bose-Einstein condensates with ferromagnetic order \cite{exp_spinor,Ueda}. Like chiral domain walls in one dimension \cite{Kim-PRB2015} and baby skyrmions in two-dimensional magnets \cite{Ochoa-PRB2016}, suitable spin-transfer torques at the interface bias the injection of these topological excitations into the frustrated magnet, which diffuse over the bulk as stable magnetic textures carrying quanta of topological charge. Robustness against structural distortions and moderate external perturbations \cite{FN2}, along with their particle-like behavior, make skyrmions attractive from the technological standpoint due to their potential use as building blocks for information and energy storage \cite{Inf-Stor,YT-PRL2018}. It is also worth remarking that frustrated magnets provide an additional condensed matter realization of these topological defects, which were originally proposed in low-energy chiral effective descriptions of QCD \cite{Skyrme-NucPhys62,Witten} and also appear in cosmology \cite{Benson-NucPhysB1993} and string theory \cite{Sakai_Sugimoto-PTP2005}.

\begin{figure}
\begin{center}
\includegraphics[width=1.0\columnwidth]{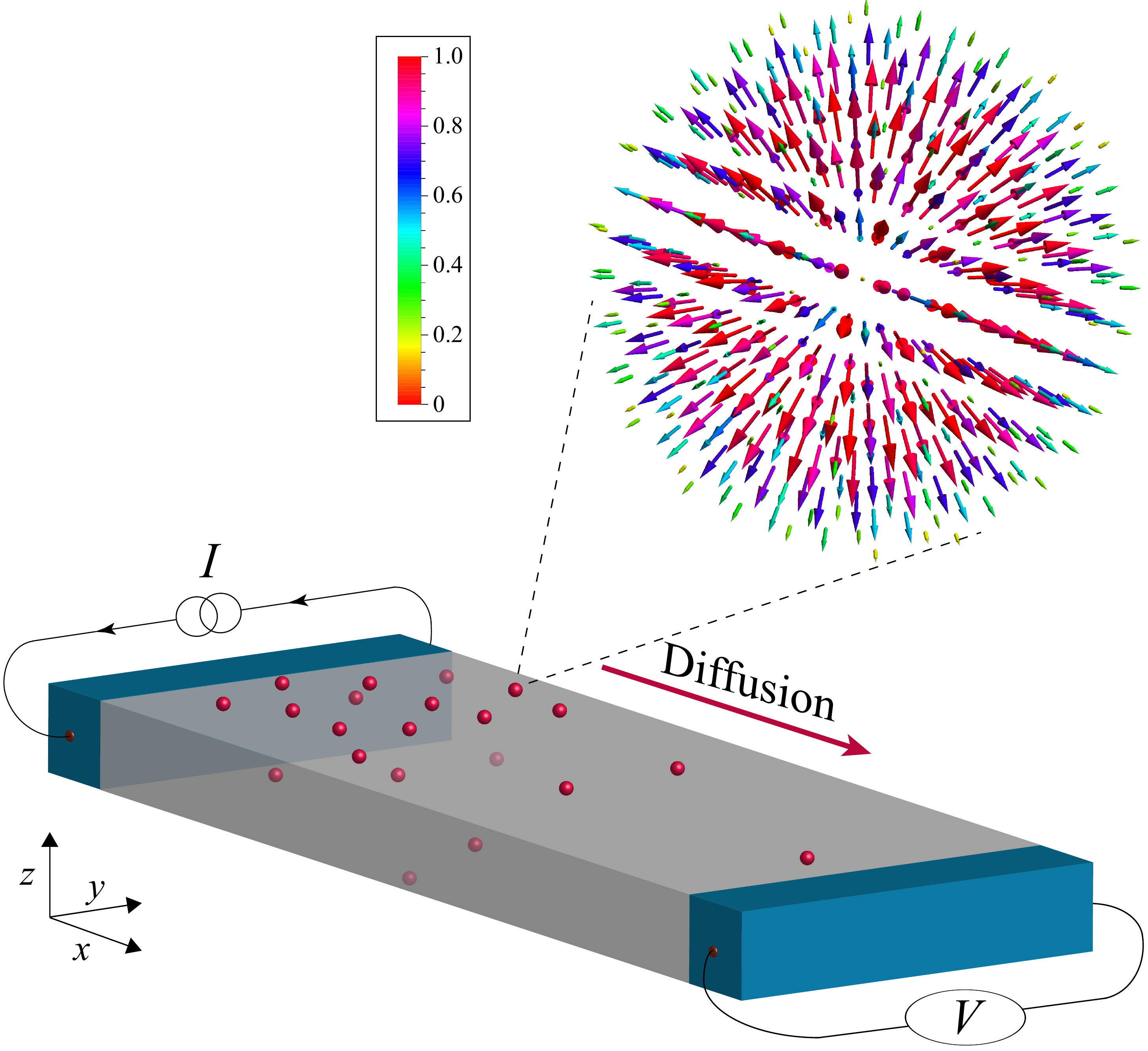}
\caption{Two-terminal geometry for the electrical injection and detection of skyrmions in frustrated magnets. Skyrmion charge is pumped into the system by the electric current flowing in the left terminal, which diffuses as stable solitons over the bulk of the magnet. The flow of topological charge across the right interface sustains a voltage signal in the second terminal via the reciprocal pumping effect. (Inset) Imaginary component of the {\it versor} parametrization of the rigid hard cut-off ansatz for skyrmions, $\mathbf{q}=\big(\hspace{-0.05cm}\cos(f(\tilde{r})/2),\sin(f(\tilde{r})/2)\bm{\hat{e}}_{r}\big)$ (see main text for details). Length and color of the arrows correspond to the magnitude of the vector field.}
\vspace{-0.5cm} 
\label{Fig1}
\end{center}
\end{figure}

In this Letter, we construct a hydrodynamic theory for skyrmions in the (electrically insulating) bulk, complemented with spin-transfer physics at the interfaces with adjacent heavy-metal contacts. Figure~\ref{Fig1} depicts the device (open) geometry with lateral terminals usually utilized in nonlocal transport measurements. For suitable crystal symmetries (Rashba-like systems), magnetic torques pump skyrmion charge into the frustrated magnet, whose diffusion over the bulk and subsequent flow across the right interface sustains a pumping electromotive force in the second terminal. The resultant drag of spin current is positive and thermally activated, in sharp contrast to the case of spin superfluid transport.

{\it Hydrodynamic theory.}|The order-parameter manifold, SO(3), is topologically equivalent to the four-dimensional unit hypersphere with antipodal points identified. 
Unit-norm quaternions (so-called versors), $\mathbf{q}=(w,\bm{v})$, provide a convenient parametrization of rotation matrices \cite{SM}: the three-dimensional vector $\bm{v}$ lies along the rotation axis, whereas the first component $w$ parametrizes the rotation angle. Skyrmions are topological defects associated with the nontrivial classes of the homotopy group $\pi_{3}\big(\textrm{SO}(3)\big)=\mathbb{Z}$, which are labeled by an integer index referred to as the skyrmion charge. The latter is the multi-dimensional analog of the winding number in 1+1 dimensions and admits the following simple expression in terms of versors:
\begin{equation}
\label{eq9}
\mathcal{Q}=\int\,d^{3}\vec{r}\,j^{0}, \hspace{0.4cm} j^{0}=\frac{\epsilon^{klm}}{12\pi^{2}}\det\left[\mathbf{q},\partial_{k}\mathbf{q},\partial_{l}\mathbf{q},\partial_{m}\mathbf{q}\right],
\end{equation}
where $k,l,m\in\{x,y,z\}$ are spatial indices, $\epsilon^{\alpha\beta\ldots\mu}$ is the Levi-Civita symbol, and $\det[\cdot,\cdot,\cdot,\cdot]$ denotes the determinant of a $4\times4$ matrix formed by versors arranged as column vectors. Our choice of prefactor ensures the normalization to unity of the skyrmion charge when the mapping $\mathbf{q}: S^{3}\rightarrow S^3$ wraps the target space once.

Formulation of a hydrodynamic theory for skyrmions requires the stability of these textures, which in turn yields the local conservation of their charge. In this regard, additional quartic terms (in the derivatives of the order parameter) in the effective action \eqref{eq1}, which may have a dipolar/exchange origin in real systems, preclude the collapse of skyrmions into atomic-size defects \cite{Skyrme-NucPhys62}. We will assume this scenario in what follows and utilize the rigid hard cut-off ansatz for stable skyrmions as a simple solution that suffices to estimate the transport coefficients of our theory \cite{FN3}:
\begin{equation}
\label{eq35}
\hat{R}(\vec{r}\,)=\exp\left[f(\tilde{r})\,\bm{\hat{e}}_{\tilde{r}}\cdot\hat{\bm{L}}\right],
\end{equation} 
where $[\hat{L}_{\alpha}]_{\beta\gamma}=-i\epsilon_{\alpha\beta\gamma}$ represent the generators of SO(3), $\tilde{r}=|\vec{r}-\vec{\mathfrak{R}}|$, and $\bm{\hat{e}}_{\tilde{r}}=(\vec{r}-\vec{\mathfrak{R}})/\tilde{r}$ is the unit radial vector from the center $\vec{\mathfrak{R}}$ of the skymion. Here, $f(\tilde{r})=2\pi(1-\tilde{r}/R_{\star})\Theta(R_{\star}-\tilde{r})$, $\Theta(x)$ denotes the Heaviside theta function and the skyrmion radius reads $R_{\star}=\xi(\mathcal{A}/\mathcal{A}_{4})^{1/2}$, where $\xi$ is a dimensionless prefactor and $\mathcal{A}_{4}$ is the strength of the fourth-order term \cite{FN4}. Figure \ref{Fig1} also depicts the vector field $\bm{v}(\vec{r}\,)=\sin(f(\tilde{r})/2)\bm{\hat{e}}_{r}$ associated with the versor parametrization of the aforementioned ansatz, Eq. \eqref{eq35}, whose skyrmion charge is $\mathcal{Q}=1$.

Topological invariance (i.e. \textit{global} conservation) of the skyrmion charge translates in our case into a \textit{local} conservation law embodied in a continuity equation. More specifically, we can cast the skyrmion charge density as the time component of a topological 4-current defined per
\begin{equation}
\label{eq10}
j^{\mu}=\frac{1}{12\pi^{2}}\epsilon^{\mu\mu_{1}\mu_{2}\mu_{3}}\det\left[\mathbf{q},\partial_{\mu_{1}}\mathbf{q},\partial_{\mu_{2}}\mathbf{q},\partial_{\mu_{3}}\mathbf{q}\right],
\end{equation}
which satisfies the continuity equation $\partial_{\mu}j^{\mu}=0$. Here, $\mu,\mu_{1,2,3}\hspace{-0.06cm}\in\hspace{-0.06cm}\{t,x,y,z\}$ denote spatio-temporal indices. The components of the associated topological flux read 
\begin{equation}
\label{eq13}
j^{k}=\frac{1}{32\pi^2}\epsilon^{klm}\,\bm{\omega}\bm{\cdot}\left(\bm{\Omega}_l\bm{\times}\bm{\Omega}_m\right),
\end{equation}
in terms of the angular velocity of the order parameter, $\bm{\omega}\equiv i\,\text{Tr}\big[\hat{R}^T\bm{\hat{L}}\,\partial_t\hat{R}\big]/2$, and the (spin) vectors $\bm{\Omega}_l\equiv i\,\text{Tr}\big[\hat{R}^T\bm{\hat{L}}\,\partial_l\hat{R}\big]/2$ describing the spatial variations of the collective spin rotation that defines the instantaneous state of the magnet \cite{OZ-PRB2018}. Note that, in the versor parametrization, these quantities can be recast as the Hamilton product $2\,\partial_{\mu}\mathbf{q}\wedge\mathbf{q}^{*}$ of the derivatives of the quaternion and its adjoint $\mathbf{q}^{*}$ \cite{SM}. Similarly, the skyrmion charge density takes the form
\begin{equation}
\label{eq15}
j^{0}=\frac{1}{16\pi^2}\,\bm{\Omega}_{z}\bm{\cdot}\left(\bm{\Omega}_{x}\bm{\times}\bm{\Omega}_{y}\right).
\end{equation}
It is worth noting here that, contrary to the case of baby skyrmions, the topological charge is even under time-reversal symmetry \cite{SM}. 

{\it Spin-transfer torques and electromotive forces}.|In general, for the device geometries considered, see Fig. \ref{Fig1}, the magnet is subject to spin-exchange and spin-orbit coupling with adjacent heavy-metal contacts. Here we construct magnetic torques compatible with the symmetries of the bulk material and assume that they also operate at the interface without invoking any reduced symmetry; thus, for our purposes, interfaces serve just as a medium for the charge current to flow. Our primary interest is to identify those magnetic torques $\bm{\tau}=\partial_{t}\bm{m}$ working in favor of the nucleation of topological charge, namely, those driving the skyrmion motion along the longitudinal direction ($x$ axis) and induced by a charge current density $\vec{J}$ flowing along the transverse direction ($y$ axis). These torques are only effective in a volume of width $\lambda$ (along $x$) in contact with the metal, where this distance characterizes the spatial extension of the proximity effect between the metal contact and the insulating magnet.

The work (per unit of volume and time) performed by the magnetic torque can be cast as $P=\bm{\tau}\bm{\cdot}\bm{\omega}$. In uniaxial magnets, where mirror reflection symmetry is broken along the $n$-fold principal axis (Rashba-like systems), this power density can be written phenomenologically as the scalar
\begin{align}
\label{eq22}
P=\frac{\hbar}{2e}\,\vec{\zeta}\cdot\big(\vec{j}\times\vec{J}\,\big),
\end{align}
where $\vec{\zeta}$ is a vector (with units of length) along the principal axis. The spin-transfer torque providing such a work is given by 
\begin{equation}
\label{eq25}
\bm{\tau}=\frac{\hbar}{32e\pi^{2}}\big(\vec{J}\cdot\vec{\bm{\Omega}}\big)\bm{\times}\big(\vec{\zeta}\cdot\vec{\bm{\Omega}}\big),
\end{equation}
which is a quadratic expression in the spatial derivatives of the order parameter. Spin torques of the form $\propto\vec{J}\cdot\vec{\bm{\Omega}}$ do not couple to the topological flux $\vec{j}$ and are therefore disregarded \cite{FN6}. We note in passing that the torque in Eq.~\eqref{eq25} breaks the macroscopic time-reversal symmetry of the spin continuity equation \cite{OZ-PRB2018}, i.e. the magnetic torque able to pump skyrmions into the bulk of the frustrated magnet is dissipative.

Skyrmion diffusion over the magnet yields a pumping electromotive force in the second terminal, whose expression can be obtained by invoking Onsager reciprocity. Currents and thermodynamic forces are related by the following matrix of linear-response coefficients:
\begin{align}
\label{eq30}
\left(\begin{array}{c}
\partial_{t}\hat{R}\\
\partial_t\bm{m}\\
\vec{J}
\end{array}\right)=\left(\begin{array}{cccc}
\cdot\star\cdot &\cdot\star\cdot  & & \cdot\star\cdot  \\
\cdot\star\cdot & \gamma\,\chi\,\bm{B}\bm{\times} & &\hat{L}_{\textrm{sq}} \\
\cdot\star\cdot & \hat{L}_{\textrm{qs}} & & \hat{\vartheta}
\end{array}\right)
\left(\begin{array}{c}
\hat{f}_{\hat{R}}\\
\bm{f}_{\bm{m}}\\
\vec{E}
\end{array}\right),
\end{align}
where $\hat{f}_{\hat{R}}\equiv-\delta\mathcal{F}/\delta\hat{R}$ and $\bm{f}_{\bm{m}}\equiv-\delta\mathcal{F}/\delta\bm{m}=-\bm{m}/\chi+\gamma\bm{B}$ $=-\bm{\omega}$ are the thermodynamic forces conjugate to the order parameter and the nonequilibrium spin density, respectively, and $\vec{E}$ represents the electromotive force. For our construction, we only need to focus on the charge and spin sectors, which are related by $\hat{L}_{\textrm{sq}}$, $\hat{L}_{\textrm{qs}}$; furthermore, $\bm{B}$ denotes an external magnetic field, $\gamma$ is the gyromagnetic ratio, and $\hat{\vartheta}$ is the conductivity tensor that we assume symmetric (i.e., purely dissipative). Note that it is not obvious whether Onsager reciprocal relations can be applied to the order-parameter sector, because the SO(3) matrices $\hat{R}$ are defined with respect to the initial (mutual equilibrium) spin configuration defining a free-energy basin, and microscopic time-reversal symmetry relates different (and possibly disconnected) basins \cite{SM}. However, the nonequilibrium spin density $\bm{m}$ does not depend on the initial configuration and, therefore, the situation for the spin-charge sectors is analogous to that of bipartite antiferromagnets \cite{Hals-PRL2011}. For the torque in Eq.~\eqref{eq25} we have
\begin{align}
\label{eq31b}
\big[\hat{L}_{\textrm{sq}}\big]_{\alpha i}&=\frac{\hbar}{32\pi^2 e}\,\epsilon_{\alpha\beta\gamma}\,\vartheta_{ij}\,\zeta_k\,\Omega_{j\beta}\,\Omega_{k\gamma},
\end{align}
and, since the off-diagonal blocks are related by the reciprocal relation $\hat{L}_{\textrm{qs}}=-\hat{L}_{\textrm{sq}}^T$, the pumping electromotive force $\vec{\mathcal{E}}=\hat{\vartheta}^{-1}\hat{L}_{\textrm{qs}}\bm{f}_{\bm{m}}$ generated in the right terminal becomes: 
\begin{align}
\label{eq34b}
\displaystyle \vec{\mathcal{E}}&=\frac{\hbar}{32 e\pi^2}\,\bm{\omega}\bm{\cdot}\left[\vec{\bm{\Omega}}\bm{\times}\left(\vec{\zeta}\cdot\vec{\bm{\Omega}}\right)\right]=\frac{\hbar}{2e}\,\vec{\zeta}\times\vec{j}.
\end{align}

{\it Skyrmion diffusion and spin drag.}|Dynamics of the soft modes (center of mass) describing stable skyrmions obey the Thiele equation
\begin{equation}
\label{eq36}
\mathcal{M}\,\ddot{\vec{\mathfrak{R}}}+\Gamma\,\dot{\vec{\mathfrak{R}}}=\vec{\mathfrak{f}},
\end{equation}
where $\mathcal{M}=\frac{16\pi}{9}(\pi^{2}+3)\chi R_{\star}$ is the skyrmion inertia and $\vec{\mathfrak{f}}=-\delta\mathcal{F}/\delta\vec{\mathfrak{R}}$ represents the thermodynamic force conjugate to the skyrmion center \cite{SM}. The friction coefficient $\Gamma=\alpha s \mathcal{M}/\chi$ is proportional to the Gilbert damping constant $\alpha$ parametrizing losses due to dissipative processes in the bulk \cite{Gilbert}, where $s\approx\hbar S/a^3$, $S$ is the length of the microscopic spin operators and $a$ denotes the lattice spacing.

Local (quasi-)equilibrium within a free-energy basin along with translational invariance in the bulk yields Fick's law for the topological flux:
\begin{equation}
\label{eq37}
\vec{j}=-D\,\vec{\nabla} j^{0},
\end{equation}
where the diffusion coefficient is related to the friction coefficient via the Einstein-Smoluchowski relation, $D=k_{\textrm{B}}T/\Gamma$. Hereafter, we assume that the current is injected into the frustrated magnet from the left contact in the two-terminal geometry depicted in Fig.~\ref{Fig1}. We also assume translational invariance along the transverse directions (i.e. the $yz$ plane). The latter, combined with the continuity equation for the topological 4-current, yields the conservation of the longitudinal bulk skyrmion current in the steady state. It reads $j^{x}_{\textrm{bulk}}=D(j^{0}_{L}-j^{0}_{R})/L_{t}$, with $j^{0}_{L/R}$ and $L_{t}$ being the skyrmion charge density at the left/right terminals and the distance between them, respectively.

The topological current at the boundaries of the magnet can be cast as
\begin{subequations}
\begin{align}
\label{eq38a}
j^{x}_{L}&=\frac{\gamma_{L}(T)\hbar\zeta\lambda J_{L}}{ek_{B}T}-\bar{\gamma}_{L}(T)j^{0}_{L},\\
\label{eq38b}
j^{x}_{R}&=\bar{\gamma}_{R}(T)j^{0}_{R},
\end{align}
\end{subequations}
where $\gamma_L(T)=\nu(T)e^{-E_{\textrm{sky}}/k_BT}$ is the equilibrium-nucleation rate of skyrmions at the left interface, $\nu(T)$ and $E_{\textrm{sky}}$ denote the attempt frequency and the skyrmion energy, respectively \cite{FN7}, and $\bar{\gamma}_{L,R}(T)$ represents the skyrmion annihilation rates per unit density \cite{Hanggi-RMP1990}. The electrical bias in the left terminal favors the nucleation of skyrmions with positive topological charge by lowering the energy barrier in an amount equal to the work carried out by the magnetic torque in Eq.~\eqref{eq25}; the expression in Eq.~\eqref{eq38a} corresponds to the leading order in the external bias \cite{Ochoa-PRB2016}. Continuity of the topological flux sets the steady state, characterized, in linear response, by the drag resistivity
\begin{equation}
\label{eq39}
\varrho_{\textrm{drag}}=\frac{\lambda R_{Q}^{2}}{R_{\textrm{bulk}}+R_{L}+R_{R}},
\end{equation}
defined per the ratio of the detected voltage per unit length to the injected charge current density. Here, $R_{Q}=h/2e^{2}\simeq 12.9$ k$\Omega$ is the quantum of resistance and $R_{\textrm{bulk}}$, $R_{L/R}$ denote the drag resistances of the bulk and interfaces of the frustrated magnet, respectively,
\begin{equation}
\label{eq40}
R_{\textrm{bulk}}=\frac{2\pi^{2}\Gamma L_{t}}{e^{2}\zeta^{2}j^{0}_{\textrm{eq}}} ,\hspace{0.4cm}R_{L/R}=\frac{2\pi^{2}k_{\textrm{B}}T}{e^{2}\zeta^{2}\gamma_{L/R}(T)},
\end{equation}
where $j^{0}_{\textrm{eq}}=\gamma_{L,R}(T)/\bar{\gamma}_{L,R}(T)=\rho_{0} e^{-E_{\textrm{sky}}/k_{\textrm{B}}T}$ is the skyrmion density at equilibrium.

\begin{figure}
\begin{center}
\includegraphics[width=1.03\columnwidth]{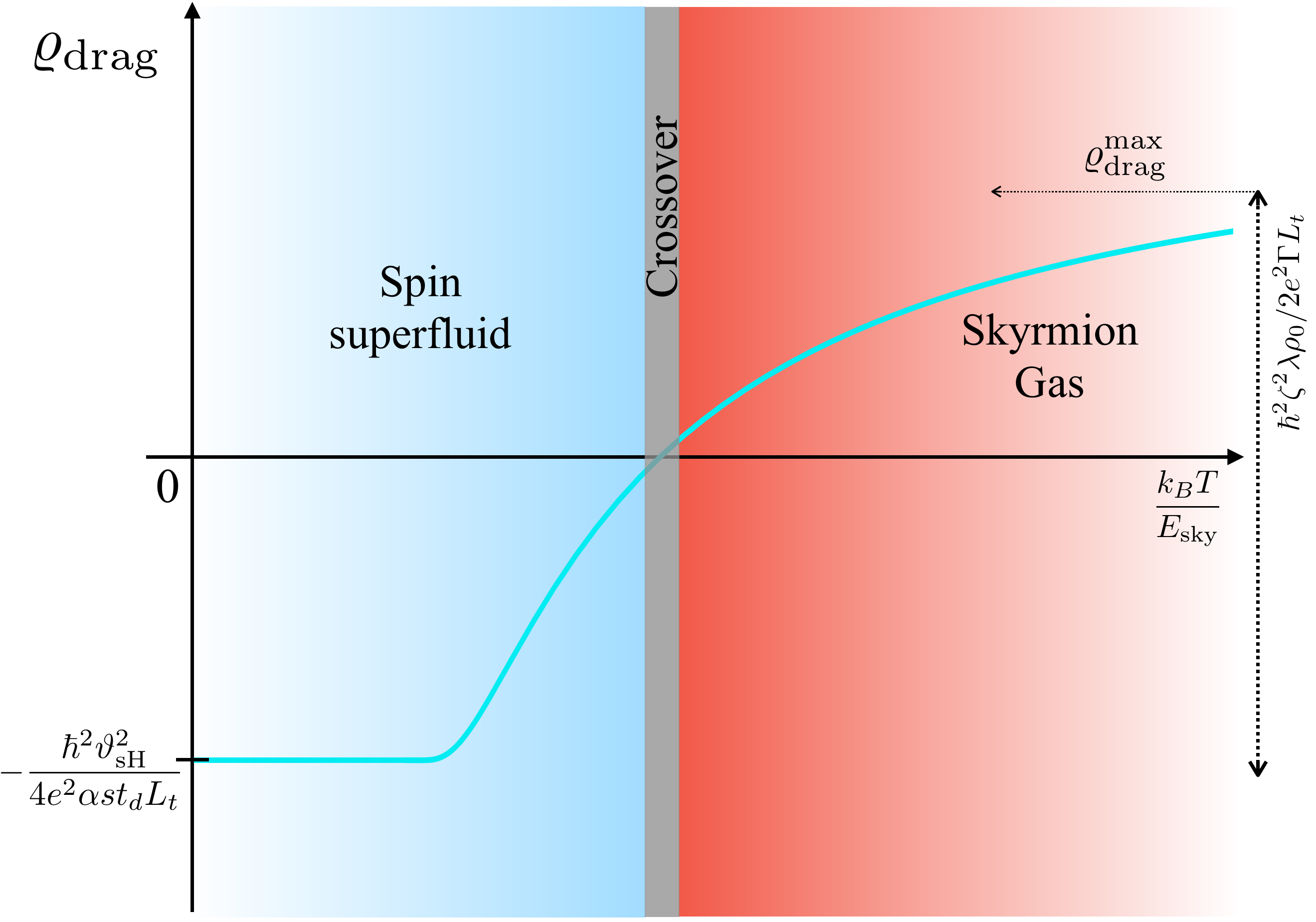}
\caption{Sketch of the thermal dependence of the total drag resistivity $\varrho_{\textrm{drag}}=\varrho_{\textrm{drag}}^{\textrm{SF}}+\varrho_{\textrm{drag}}^{\textrm{sky}}$ for frustrated magnets with weak easy-plane anisotropies.
}
\vspace{-0.5cm} 
\label{Fig2}
\end{center}
\end{figure}

{\it Discussion.}|The channel for spin transport rooted in the diffusion of skyrmion charge becomes suppressed in the low-temperature regime, as the proliferation of skyrmions in the bulk of the magnet dies out with probability $\propto e^{-E_{\textrm{sky}}/k_{\textrm{B}}T}$. The frustrated magnet, however, sustains stable spin supercurrents at low temperatures in the presence of additional easy-plane anisotropies; this coherent transport of spin may be driven by nonequilibrium spin accumulations at the left interface, which are induced by the charge current flowing within the first terminal via the spin Hall effect \cite{OZ-PRB2018}. Furthermore, in the absence of topological singularities in the SO(3) order parameter (namely, $\mathbb{Z}_{2}$ vortices) degradation of the spin superflow only occurs via thermally-activated phase slips in the form of $4\pi$-vortex lines \cite{OZ-PRB2018}. In that regard, we can show through the analog of the Mermin-Ho relation \cite{Mermin-Ho,SM}, $\vec{\nabla}\times\vec{J}_{\alpha}=-(\mathcal{A}/2)\epsilon_{\alpha\beta\gamma}\vec{\Omega}_{\alpha}\times\vec{\Omega}_{\beta}$ (here, $\vec{J}_{\alpha}=-\mathcal{A}\,\vec{\Omega}_{\alpha}$ is the $\alpha$-component of the spin supercurrent), that skyrmions crossing streamlines in a planar section of the magnet do not contribute to the generation of phase slips in the superfluid \cite{FN8}. Therefore, in magnetically frustrated systems with weak easy-plane anisotropies, we expect to observe a smooth crossover from a spin superfluid to a \textit{skyrmion conductor} driven by temperature, as depicted in Fig.~\ref{Fig2}. For a large separation between terminals, $L_{t}\gg1/\Gamma \gamma_{L,R}, \hbar g_{L,R}/4\pi\alpha s$ ($g_{L,R}$ are the effective interfacial conductances), the drag coefficients for both transport channels reduce to
\begin{equation}
\label{eq41}
\hspace{-0.2cm}\varrho_{\textrm{drag}}^{\textrm{sky}}=\left(\hspace{-0.05cm}\frac{\hbar}{e}\hspace{-0.05cm}\right)^{2}\hspace{-0.05cm}\frac{\zeta^{2}\lambda\,j^{0}_{\textrm{eq}}}{2\Gamma L_{t}}, \hspace{0.2cm}\varrho_{\textrm{drag}}^{\textrm{SF}}=-\left(\hspace{-0.05cm}\frac{\hbar}{2e}\hspace{-0.05cm}\right)^{2}\hspace{-0.05cm}\frac{\vartheta_{\textrm{sH}}^{2}}{\alpha s t_{d} L_{t}},
\end{equation}
where $\vartheta_{\textrm{sH}}$ and $t_{d}$ denote the spin Hall angle in the metal contacts and the thickness of the detector strip, respectively. Note the algebraical decay $\varrho_{\textrm{drag}}^{\textrm{sky,SF}}\propto 1/L_{t}$ and the opposite sign of the drag resistivities in these two spin-transport channels. The latter can be intuitively understood as the manifestation of the different symmetries under time reversal of the flavors encoding the information and dragging of the electrical signal: while in the case of the superfluid this is just the spin flow ascribed to coherent precession, in the case of the skyrmion conductor the signal is mediated by the flux of the associated topological charge, which is even under time reversal. As a final remark, we note in passing that, remarkably, skyrmions do generate hopfions through the fibration $S^{3}\rightarrow S^{2}$ described by a given element of the internal spin frame \cite{FN9}. 

In conclusion, we have established the hydrodynamic equations governing the diffusion of skyrmion charge within the bulk of frustrated magnetic insulators. Interfacial spin-transfer torques inject topological charge into the system, whose steady flow sustains a spin drag signal between the metallic terminals. The algebraic decay of the drag coefficient over long distances manifests the topological robustness of Shankar skyrmions in the SO(3) order parameter. We also remark that $S^{2}$ hopfions could be pumped into the frustrated magnet by suitable spin-transfer torques, therefore giving rise to a third channel for low-dissipation spin transport. The program developed in this Letter can, in principle, be extended to $S^{2}$ hopfions, with the caveat that the Hopf charge density is nonlocal in the order parameter \cite{Whitehead-PNAS1947} and that it is unclear whether these topological excitations are stable within Skyrme-like models \cite{Foster-JPhysA2017}.

\begin{acknowledgments}

This work has been supported by NSF under Grant No. DMR-1742928.

R.Z. and H.O. contributed equally to this work.

\end{acknowledgments}

\onecolumngrid
 
 \newpage
 
\hspace{0.1cm}
\section*{\Large Supplemental Material}
\hspace{0.5cm}

\section{Versor parametrization}
\label{SM:A}

In this Section we show that versors (i.e., unit-norm quaternions) provide a convenient parametrization of rotation matrices. To begin with, note that SU(2) is the universal (double) covering of SO(3) and is also isomorphic to the unit hypersphere in $\mathbb{R}^{4}$. The latter means that we can represent a generic $2\times2$ complex matrix $\hat{\mathcal{U}}$ by means of a 4-component vector $\mathbf{q}=(w,\bm{v})$:
\begin{align}
\label{eq:def_quaternion}
\hat{\mathcal{U}}=w\hat{1}-i \bm{v}\cdot\bm{\sigma}\equiv w\hat{1}-iv_{x}\hat{\sigma}_x-iv_{y}\hat{\sigma}_y-iv_{z}\hat{\sigma}_z,
\end{align}
where $\bm{\sigma}=(\hat{\sigma}_{x},\hat{\sigma}_{y},\hat{\sigma}_{z})$ is the vector of Pauli matrices and $\bm{v}=(v_{x},v_{y},v_{z})$ denotes the vector (imaginary) part of the quaternion. The above isomorphism sets a one-to-one correspondence between unitary operators, $\det\,\hat{\mathcal{U}}=1$, and unit-norm quaternions, $w^2+\bm{v}^2=1$. The SO(3) matrix $\hat{R}$ associated with $\hat{\mathcal{U}}\in\textrm{SU(2)}$ reads
\begin{align}
\label{eq:quaternion_parametrization}
\hat{R}_{\alpha\beta}=\left(1-2\left|\bm{v}\right|^2\right)\delta_{\alpha\beta}+2\,v_{\alpha}v_{\beta}-2\,\varepsilon_{\alpha\beta\gamma}\,w\,v_{\gamma}.
\end{align}
Since $\mathbf{q}$ and $-\mathbf{q}$ parametrize the same rotation $\hat{R}$, we conclude that SO$(3)\cong \mathbb{R}\textrm{P}^{3}$, namely the group of proper rotations corresponds to the hypersphere $S^{3}$ with antipodal points being identified. In this parametrization of rotations, $\bm{v}$ lies along the rotation axis and the first component $w$ parametrizes the rotation angle.

The set $\{\hat{1},-i\hat{\sigma}_x,-i\hat{\sigma}_y,-i\hat{\sigma}_z\}$ defines the basis of quaternions as a real vector space, where addition and multiplication by scalars is as in $\mathbb{R}^4$. The Hamilton product
\begin{align}
\label{eq:Hamilton}
\mathbf{q}_1\wedge\mathbf{q}_2\equiv\left(w_1w_2-\bm{v}_1\cdot\bm{v}_2,w_1\bm{v}_2+w_2\bm{v}_1+\bm{v}_1\times\bm{v}_2\right),
\end{align}
is inferred directly from the algebra of Pauli matrices. The adjoint of $\mathbf{q}=(w,\bm{v})$ is $\mathbf{q}^{*}=(w,-\bm{v})$, so that the norm $\sqrt{\mathbf{q}^{*}\wedge\mathbf{q}}$ ($=1$ in the case of versors) is a real number. Note that the Hamilton product provides a convenient representation of the usual matrix product in SO(3), since $\mathbf{q}^{*}$ corresponds to $\hat{R}^T$ and $\hat{R}_{1}\cdot\hat{R}_{2}$ corresponds to $\mathbf{q}_{1}\wedge\mathbf{q}_{2}$.

Finally, the O(4) nonlinear $\sigma-$model takes the following simple form
\begin{equation}
\label{eq:nlsigma}
\mathcal{L}=2\int\,d^{3}\vec{r}\left(\chi\,\partial_t\mathbf{q}^{*}\wedge\partial_t\mathbf{q}-\mathcal{A}\,\partial_i\mathbf{q}^{*}\wedge\partial_i\mathbf{q}\right),
\end{equation}
in terms of versors. A simple spin-wave analysis of this Lagrangian yields, akin to N\'{e}el antiferromagnets, three independent linear dispersion relations characterized by the sound velocity $c=\sqrt{\mathcal{A}/\chi}$. 

\section{Versors under parity and time-reversal symmetries}
\label{SM:B}

The order-parameter manifold of magnetic systems with frustrated interactions dominated by exchange is generically built upon applying SO(3) rotations to a given ground state $G$, which corresponds to a classical solution (a minimum) of the free-energy landscape \cite{Halperin-PRB77,Dombre-PRB89,OZ-PRB2018}. These rotations connect physically distinguishable spin configurations with the same energy. Nonequilibrium deviations within the free-energy basin are described by smoothly varying (in space and time) elements of SO(3) in this approach.

Let $\hat{\textrm{P}}$ and $\hat{\textrm{T}}$ be the operators (in spin space) corresponding to the representations of parity and time-reversal symmetry operations, respectively. Note that the action of these symmetries on the ground state $|G\rangle$ leads to isoenergetic states $|G'\rangle$ that belong in general to other energy basins. The spin rotation operator $\hat{\mathcal{U}}$ acting on the whole set of spins is the direct sum of irreducible representations of SU(2) acting on individual spins $\mathbf{S}_i$ ($i$ labels here the spatial position). 
The identities $|G'\rangle\equiv\hat{\textrm{T}}\hat{\mathcal{U}}|G\rangle=\hat{\mathcal{U}}\hat{\textrm{T}}|G\rangle$ and $|G''\rangle\equiv\hat{\textrm{P}}\hat{\mathcal{U}}|G\rangle=\hat{\mathcal{U}}\hat{\textrm{P}}|G\rangle$ follow from 
\begin{subequations}
\begin{align}
\label{eq7a}
& \hat{\textrm{T}}\,\mathbf{S}_i\,\hat{\textrm{T}}^{-1}=-\mathbf{S}_i,\\
\label{eq7b}
& \hat{\textrm{P}}\,\mathbf{S}_i\,\hat{\textrm{P}}^{-1}=\mathbf{S}_i,
\end{align}
\end{subequations}
so that $\hat{\textrm{P}}\,\hat{\mathcal{U}}\,\hat{\textrm{P}}^{-1}=\hat{\mathcal{U}}$ and $\hat{\textrm{T}}\,\hat{\mathcal{U}}\,\hat{\textrm{T}}^{-1}=\hat{\mathcal{U}}$. With account of Eq. \eqref{eq:def_quaternion}, we have the identities
\begin{subequations}
\begin{align}
\label{eq8a}
&\bullet\hspace{0.5cm}\hat{\textrm{P}}\,\hat{\mathcal{U}}\,\hat{\textrm{P}}^{-1}=\hat{\textrm{P}}w\hat{\textrm{P}}^{-1}-i\hat{\textrm{P}}\bm{v}\hat{\textrm{P}}^{-1}\cdot\hat{\textrm{P}}\bm{\sigma}\hat{\textrm{P}}^{-1}=w-i\bm{v}\cdot\bm{\sigma}=\hat{\mathcal{U}}\Longrightarrow \mathbf{q}\stackrel{\hat{\textrm{P}}}{\mapsto}\mathbf{q},\\
\label{eq8b}
&\bullet\hspace{0.5cm}\hat{\textrm{T}}\,\hat{\mathcal{U}}\,\hat{\textrm{T}}^{-1}=\hat{\textrm{T}}w\hat{\textrm{T}}^{-1}+i\hat{\textrm{T}}\bm{v}\hat{\textrm{T}}^{-1}\cdot\hat{\textrm{T}}\bm{\sigma}\hat{\textrm{T}}^{-1}=w-i\bm{v}\cdot\bm{\sigma}=\hat{\mathcal{U}}\Longrightarrow \mathbf{q}\stackrel{\hat{\textrm{T}}}{\mapsto}\mathbf{q},
\end{align}\end{subequations}

Thus, the quaternions that parametrize SO(3) rotations (with respect to the new basin) remain invariant under the inversion operations. 

\section{Mermin-Ho relation}
\label{SM:C}

We first introduce the fields $\bm{\Omega}_{\mu}=i\,\text{Tr}\big[\hat{R}^T\bm{\hat{L}}\,\partial_{\mu}\hat{R}\big]/2$, which describe time ($\mu=t$) and spatial ($\mu=x,y,z$) variations of the collective spin rotation defining the instantaneous state of the magnet,
\begin{equation}
\label{eq:Omega}
i\,\partial_{\mu} \,\hat{\mathcal{U}}\left(t,\vec{r}\right)=\big(\bm{\Omega}_{\mu}\left(t,\vec{r}\right)\cdot\mathbf{\hat{S}}\big)\,\hat{\mathcal{U}}\left(t,\vec{r}\right).
\end{equation}
Here, $[\hat{L}_{\alpha}]_{\beta\gamma}=-i\epsilon_{\alpha\beta\gamma}$ are the generators of SO(3) and $\epsilon_{\alpha\beta\gamma}$ is the Levi-Civita symbol. In particular, $\bm{\Omega}_{t}=\bm{\omega}$ is the angular velocity of the order parameter $\hat{R}$. The spin current is given by $\vec{\bm{J}}=-\mathcal{A}\,\vec{\bm{\Omega}}$, as inferred from the Euler-Lagrange equations, where $\vec{\bm{\Omega}}\equiv\bm{\Omega}_{x}\hat{e}_{x}+\bm{\Omega}_{y}\hat{e}_{y}+\bm{\Omega}_{z}\hat{e}_{z}$ \cite{OZ-PRB2018}.  With account of the versor parametrization, Eq. \eqref{eq:quaternion_parametrization}, we obtain the identity
\begin{align}
\label{eq:Omega_quaternion2}
\bm{\Omega}_{\mu}=2w\partial_{\mu}\bm{v}-2\bm{v}\partial_{\mu}w+2\bm{v}\times\partial_{\mu}\bm{v},
\end{align}
which is just $\bm{\Omega}_{\mu}=2\,\partial_{\mu}\mathbf{q}\wedge\mathbf{q}^{*}$ as deduced from the definition of the Hamilton product, Eq.~\eqref{eq:Hamilton}. Note that the \textit{scalar} part of $\partial_{\mu}\mathbf{q}\wedge\mathbf{q}^{*}$ is identically zero, $w\partial_{\mu} w+v_{\alpha}\partial_{\mu} v_{\alpha}=0$.

The following identity holds in the absence of singularities in the order parameter:
\begin{align}
\label{eq:Mermin-Ho1}
\partial_{\mu_{1}}\bm{\Omega}_{\mu_{2}}-\partial_{\mu_{2}}\bm{\Omega}_{\mu_{1}}=\bm{\Omega}_{\mu_{1}}\bm{\times}\bm{\Omega}_{\mu_{2}}.\hspace{0.5cm}\mu_{1},\mu_{2}\in\{t,x,y,z\}
\end{align}
In terms of the $\alpha $-component of the spin current, $\vec{J}_{\alpha}=-\mathcal{A}\,\vec{\Omega}_{\alpha }$, the above equation for spatial subindices can be recast as
\begin{align}
\label{eq:Mermin-Ho2}
\vec{\nabla}\times\vec{J}_{\alpha}=-\frac{\mathcal{A}}{2}\epsilon_{\alpha \beta\gamma}\,\vec{\Omega}_{\beta}\times\vec{\Omega}_{\gamma},
\end{align}
which is analogous to the Mermin-Ho relation in $^3$He-A \cite{Mermin-Ho}. Equation~\eqref{eq:Mermin-Ho1} can be easily proved in versor notation since 
\begin{align}
\partial_{\mu_{1}}\bm{\Omega}_{\mu_{2}}-\partial_{\mu_{2}}\bm{\Omega}_{\mu_{1}}=2\,\partial_{\mu_{2}}\mathbf{q}\wedge\partial_{\mu_{1}}\mathbf{q}^{*}-2\,\partial_{\mu_{1}}\mathbf{q}\wedge\partial_{\mu_{2}}\mathbf{q}^{*}=\bm{\Omega}_{\mu_{1}}\bm{\times}\bm{\Omega}_{\mu_{2}},
\end{align}
so long as the order parameter is singe-valued and, therefore, $\partial_{\mu_{1}}\partial_{\mu_{2}}\mathbf{q}=\partial_{\mu_{2}}\partial_{\mu_{1}}\mathbf{q}$. 

The internal spin frame of reference is defined locally by the tetrad of vectors $\bm{\hat{e}}_{\alpha }=\hat{R}\cdot\hat{\bm{\alpha }}$, $\alpha=x,y,z$. By projecting Eq. \eqref{eq:Mermin-Ho1} onto these director vectors, we obtain
\begin{equation}
\label{eq:tetrad}
\bm{\hat{e}}_{\alpha }\bm{\cdot}(\bm{\Omega}_i\bm{\times}\bm{\Omega}_j)=
\bm{\hat{e}}_{\alpha }\bm{\cdot}(\partial_{i}\bm{\hat{e}}_{\alpha }\times\partial_{j}\bm{\hat{e}}_{\alpha }).
\end{equation}
Furthermore, the projection of the spin current onto the vectors $\{\bm{\hat{e}}_{\alpha }\}_{\alpha }$ defines the components of the {\it internal spin current}, namely the spin current measured in the internal spin frame of the texture: 
\begin{equation}
\vec{J}_{(\alpha)}=\bm{\hat{e}}_{\alpha }\bm{\cdot}\vec{\bm{J}}=\left[\hat{R}^{T}\hspace{-0.1cm}\cdot\vec{\bm{J}}\right]_{\alpha }.
\end{equation}
From Eq. \eqref{eq:Mermin-Ho1} we obtain the Mermin-Ho-like identity
\begin{align}
\label{eq:Mermin-Ho3}
\left[\vec{\nabla}\times\vec{J}_{(\alpha)}\right]_{k}=-\frac{\mathcal{A}}{2}\epsilon_{ijk}\,\bm{\hat{e}}_{\alpha }\bm{\cdot}(\partial_{i}\bm{\hat{e}}_{\alpha }\times\partial_{j}\bm{\hat{e}}_{\alpha }), 
\end{align}
which implies that the circulation of the $\alpha $-component of the internal spin current along a closed loop is proportional to the solid angle subtended by the surface defined by $\bm{\hat{e}}_{\alpha }$ on the planar section enclosed by the loop. Therefore, in the absence of singularities in the order parameter, the spin current can only decay in multiples of $4\pi\mathcal{A}$ because the solid angle is quantized in units of $4\pi$ (provided that $\bm{\hat{e}}_{\alpha }$ points towards the same direction far away from the phase-slip event).

\section{Collective-variable approach for skyrmions}
\label{SM:G}

Time dependence of the SO(3)-order parameter for the hard cut-off ansatz is encoded in the soft modes of the skyrmion texture, namely its center of mass: $\hat{R}(t,\vec{r})\equiv\hat{R}\big[\vec{r}-\vec{\mathfrak{R}}(t)\big]$. At the same time, the canonical momentum $\vec{\Pi}$ conjugate to $\vec{\mathfrak{R}}$ reads
\begin{equation}
\vec{\Pi}=-\int d^{3}\vec{r}\hspace{0.15cm}\bm{m}\bm{\cdot}\vec{\bm{\Omega}}.
\end{equation}
With account of the equation of motion, $\bm{m}=\chi\bm{\omega}$, and of $\partial_{t}\hat{R}\approx-\big(\dot{\vec{\mathfrak{R}}}\cdot\vec{\nabla}_{\vec{r}}\big)\hat{R}$ for rigid skyrmions, we can write the canonical momentum as $\Pi_{i}=M_{ij}\,\dot{\mathfrak{R}}_{j}$, where the inertia tensor takes the form:
\begin{equation}
M_{ij}=\chi\int d^{3}\vec{r}\hspace{0.15cm}\bm{\Omega}_{i}\cdot\bm{\Omega}_{j}=4\chi\int d^{3}\vec{r}\hspace{0.15cm}\left(\partial_{i}w\partial_{j}w+\partial_{i}\bm{v}\cdot\partial_{j}\bm{v}\right)=\mathcal{M}\,\delta_{ij}, \hspace{0.5cm}\mathcal{M}=\frac{16\pi}{9}(\pi^{2}+3)\chi R_{\star}.
\end{equation}
For the final result, we have used the ansatz given in the main text.

We model dissipation by means of the Gilbert-Rayleigh function \cite{OZ-PRB2018} 
\begin{equation}
\displaystyle \mathcal{R}\big[\hat{R}\big]=\frac{\alpha s}{2}\int d^{3}\vec{r}\,\bm{\omega}^{2}=\frac{\alpha s}{4}\int d^{3}\vec{r}\hspace{0.15cm}\textrm{Tr}\left[\partial_{t}\hat{R}^{T}\partial_{t}\hat{R}\right],
\end{equation}
which provides the dominant term in the low-frequency (compared to the microscopic exchange $J$) regime. Within the collective-variable approach it becomes
\begin{equation}
\label{F4}
\mathcal{R}\big[\hat{R}\big]=\frac{1}{2}\dot{\vec{\mathfrak{R}}}\cdot[\Gamma]\cdot\dot{\vec{\mathfrak{R}}},\hspace{0.5cm}\Gamma_{ij}=\frac{M_{ij}}{\mathcal{T}},
\end{equation}
where $\mathcal{T}=\chi/s\alpha$ represents a relaxation time. Therefore, the Euler-Lagrange equations for the skyrmion center (the so-called Thiele equation) turn out to be:
\begin{equation}
\label{F5}
\mathcal{M}\,\ddot{\vec{\mathfrak{R}}}+\frac{\mathcal{M}}{\mathcal{T}}\,\dot{\vec{\mathfrak{R}}}=\vec{\mathfrak{f}},
\end{equation}
where $\vec{\mathfrak{f}}\equiv-\delta_{\vec{\mathfrak{R}}}\mathcal{F}$ is the conservative force.


\begin{thebibliography}{100}
\bibitem{FN1} It includes, e.g., the correlated-spin-glass phase of amorphous magnets and multi-lattice antiferromagnets with geometrical frustration.
\bibitem{FN2} Skyrmions are classified according to a topological integer invariant, the so-called skyrmion charge. Continuous deformations of the spin structure do not modify this charge, so that skyrmion textures are disconnected from the ground state by an energy gap of topological origin.
\bibitem{FN3} In addition to its center $\vec{\mathfrak{R}}=(X,Y,Z)$ and radius $R$, the ansatz in Eq.~\eqref{eq35} is implicitly parametrized by two angular offsets $\theta_{0}$ (polar) and $\phi_{0}$ (azimuthal) describing the orientation of $\bm{\hat{e}}_{\tilde{r}}$ in the spin frame defined by the 3 generators of SO(3). This broken symmetry is generically restored by quantum fluctuations, provided that $\theta_{0}$ and $\phi_{0}$ are compact variables and, thus, the associated spectrum consists of discrete levels labelled by integer angular-momentum numbers. In the presence of additional free-energy terms with a quartic dependence on the derivatives of the order parameter \cite{FN4}, fluctuations of the radius around $R_{*}$ are energetically penalized. As a result, only the coordinates $\vec{\mathfrak{R}}$ represent true soft modes of these topological defects.
\bibitem{FN4} In particular, skyrmions are stabilized by the so-called Skyrme term, $S_{4}=(\mathcal{A}_{4}/8)\int\,d^{3}\vec{r}\,dt\, \textrm{Tr}\big[\partial_{\mu_{1}}\hat{R}^{T}\partial_{\mu_{2}}\hat{R}-\partial_{\mu_{2}}\hat{R}^{T}\partial_{\mu_{1}}\hat{R}\big]^{2}$, see Ref. \onlinecite{Skyrme-NucPhys62}.
\bibitem{FN6} Other broken symmetries could give rise to different possible couplings. For instance, in noncentrosymmetric magnets, the power density can be identified with the pseudoscalar $P=\frac{\hbar\delta}{2e}\vec{J}\cdot\vec{j}$, where $\delta$ is a phenomenological parameter with dimension of length. In this case, the spin-transfer torque yielding this work reads $\bm{\tau}=\frac{\hbar\delta}{32\pi^{2}e}\vec{J}\cdot\big(\vec{\nabla}\times\vec{\bm{\Omega}}\big)$. Note that this torque drives the diffusion of skyrmions along the transverse $y$ direction (not between the terminals) and also breaks macroscopic time-reversal symmetry.
\bibitem{FN8} By projecting this relation onto the element $\bm{\hat{e}}_{z}\equiv\hat{R}\cdot\bm{\hat{z}}$ of the tetrad defining the internal frame of reference, we obtain the expression $-4\pi\mathcal{A}\rho_{\textrm{sky}}^{\textrm{2D}}$ for the right-hand side, where $\rho_{\textrm{sky}}^{\textrm{2D}}=\bm{\hat{e}}_{z}\cdot(\partial_{x}\bm{\hat{e}}_{z}\times\partial_{y}\bm{\hat{e}}_{z})/4\pi$ is the baby-skyrmion charge density associated with the vector field $\bm{\hat{e}}_{z}:\mathbb{R}^{3}\mapsto S^{2}$ restricted to planar $z$-sections. As a result, in the absence of disclinations, the spin superflow can decay in multiples of $4\pi\mathcal{A}$ only, since the solid angle swept by the $z$-element of the tetrad is quantized in units of $4\pi$. The hard cut-off ansatz for skyrmions yields a total solid angle of zero (per planar section). On the other hand, coreless $4\pi$-vortex of the SO(3) order parameter in the $xy$ plane correspond to baby skyrmions of the field $\bm{\hat{e}}_{z}$.
\bibitem{FN9} Hopfions constitute the nontrivial classes of the homotopy group $\pi_{3}(S^{2})=\mathbb{Z}$. These topological textures are classified by a linking number, the so-called Hopf charge. In the case of skyrmion-projected hopfion textures, described by, e.g., the $z$-element of the internal frame of reference, $\mathbf{q}\mapsto \hat{R}[\mathbf{q}]\cdot\bm{\hat{z}}=(2v_{x}v_{z}+2wv_{y},2v_{y}v_{z}-wv_{x},1-2v_{x}^{2}-2v_{y}^{2})^{T}$, their Hopf charge corresponds to the skyrmion charge \eqref{eq9}.
\end{thebibliography}
\end{document}